\documentclass[%
 reprint,
superscriptaddress,
 amsmath,amssymb,
 aps,
]{revtex4-2}
\usepackage{amsmath}
\usepackage{lineno}
\usepackage{graphicx}
\usepackage{dcolumn}
\usepackage{bm}
\usepackage{xcolor}
\usepackage{orcidlink}
\renewcommand{\thefigure}{\arabic{figure}}
\newcommand{\orcid}[1]{\,\orcidlink{#1}}

\begin{document}

\preprint{APS/123-QED}
\title{First Optical Observation of Negative Ion Drift at Surface Pressure}

\author{F.~D.~Amaro\orcid{0000-0001-7315-0550}}
\affiliation{LIBPhys, Department of Physics, University of Coimbra, 3004-516 Coimbra, Portugal}

\author{R.~Antonietti\orcid{0009-0009-2568-8247}}
\affiliation{Dipartimento di Matematica e Fisica, Università Roma Tre, 00146 Roma, Italy}
\affiliation{Istituto Nazionale di Fisica Nucleare, Sezione di Roma Tre, 00146 Roma, Italy}

\author{E.~Baracchini\orcid{0000-0003-4686-128X}}
\affiliation{Gran Sasso Science Institute, 67100 L'Aquila, Italy}
\affiliation{Istituto Nazionale di Fisica Nucleare, Laboratori Nazionali del Gran Sasso, 67100 Assergi, Italy}

\author{L.~Benussi\orcid{0000-0002-2363-8889}}
\affiliation{Istituto Nazionale di Fisica Nucleare, Laboratori Nazionali di Frascati, 00044 Frascati, Italy}

\author{C.~Capoccia\orcid{0009-0008-5919-3130}}
\affiliation{Istituto Nazionale di Fisica Nucleare, Laboratori Nazionali di Frascati, 00044 Frascati, Italy}

\author{M.~Caponero\orcid{0000-0002-5728-3123}}
\affiliation{Istituto Nazionale di Fisica Nucleare, Laboratori Nazionali di Frascati, 00044 Frascati, Italy}
\affiliation{ENEA Centro Ricerche Frascati, 00044 Frascati, Italy}

\author{L.~G.~M.~de~Carvalho\orcid{0009-0003-5836-4771}}
\affiliation{Universidade Federal de Juiz de Fora, Faculdade de Engenharia, 36036-900 Juiz de Fora, MG, Brazil}

\author{G.~Cavoto\orcid{0000-0003-2161-918X}}
\affiliation{Dipartimento di Fisica, Sapienza Università di Roma, 00185 Roma, Italy}
\affiliation{Istituto Nazionale di Fisica Nucleare, Sezione di Roma, 00185 Roma, Italy}

\author{I.~A.~Costa\orcid{0000-0002-3064-8305}}
\affiliation{Istituto Nazionale di Fisica Nucleare, Laboratori Nazionali di Frascati, 00044 Frascati, Italy}

\author{A.~Croce}
\affiliation{Istituto Nazionale di Fisica Nucleare, Laboratori Nazionali di Frascati, 00044 Frascati, Italy}

\author{M.~D'Astolfo\orcid{0009-0000-9817-6693}}
\affiliation{Gran Sasso Science Institute, 67100 L'Aquila, Italy}
\affiliation{Istituto Nazionale di Fisica Nucleare, Laboratori Nazionali del Gran Sasso, 67100 Assergi, Italy}

\author{G.~D'Imperio\orcid{0000-0002-2945-0983}}
\affiliation{Istituto Nazionale di Fisica Nucleare, Sezione di Roma, 00185 Roma, Italy}

\author{G.~Dho\orcid{0000-0001-9454-9894}}
\affiliation{Istituto Nazionale di Fisica Nucleare, Laboratori Nazionali di Frascati, 00044 Frascati, Italy}

\author{F.~Di~Giambattista\orcid{0000-0001-7831-3961}}
\affiliation{Gran Sasso Science Institute, 67100 L'Aquila, Italy}
\affiliation{Istituto Nazionale di Fisica Nucleare, Laboratori Nazionali del Gran Sasso, 67100 Assergi, Italy}

\author{E.~Di~Marco\orcid{0000-0002-5920-2438}}
\affiliation{Istituto Nazionale di Fisica Nucleare, Sezione di Roma, 00185 Roma, Italy}

\author{J.~M.~F.~dos~Santos\orcid{0000-0002-8841-6523}}
\affiliation{LIBPhys, Department of Physics, University of Coimbra, 3004-516 Coimbra, Portugal}

\author{D.~Fiorina\orcid{0000-0002-7104-257X}}
\affiliation{Gran Sasso Science Institute, 67100 L'Aquila, Italy}
\affiliation{Istituto Nazionale di Fisica Nucleare, Laboratori Nazionali del Gran Sasso, 67100 Assergi, Italy}

\author{F.~Iacoangeli\orcid{0000-0003-0808-585}}
\affiliation{Istituto Nazionale di Fisica Nucleare, Sezione di Roma, 00185 Roma, Italy}

\author{Z.~Islam\orcid{0000-0003-4611-839X}}
\affiliation{Gran Sasso Science Institute, 67100 L'Aquila, Italy}
\affiliation{Istituto Nazionale di Fisica Nucleare, Laboratori Nazionali del Gran Sasso, 67100 Assergi, Italy}


\author{H.~P.~Lima~Jr.\orcid{0000-0001-7398-3237}}
\affiliation{Gran Sasso Science Institute, 67100 L'Aquila, Italy}
\affiliation{Istituto Nazionale di Fisica Nucleare, Laboratori Nazionali del Gran Sasso, 67100 Assergi, Italy}

\author{G.~Maccarrone\orcid{0000-0002-7234-9522}}
\affiliation{Istituto Nazionale di Fisica Nucleare, Laboratori Nazionali di Frascati, 00044 Frascati, Italy}

\author{R.~D.~P.~Mano\orcid{0000-0003-2920-7067}}
\affiliation{LIBPhys, Department of Physics, University of Coimbra, 3004-516 Coimbra, Portugal}

\author{D.~J.~G.~Marques\orcid{0000-0002-0013-6341}}
\affiliation{Gran Sasso Science Institute, 67100 L'Aquila, Italy}
\affiliation{Istituto Nazionale di Fisica Nucleare, Laboratori Nazionali del Gran Sasso, 67100 Assergi, Italy}

\author{G.~Mazzitelli\orcid{0000-0003-2830-4359}}
\affiliation{Istituto Nazionale di Fisica Nucleare, Laboratori Nazionali di Frascati, 00044 Frascati, Italy}

\author{P.~Meloni\orcid{0009-0001-7634-370X}}
\affiliation{Dipartimento di Matematica e Fisica, Università Roma Tre, 00146 Roma, Italy}
\affiliation{Istituto Nazionale di Fisica Nucleare, Sezione di Roma Tre, 00146 Roma, Italy}

\author{A.~Messina\orcid{0000-0003-1195-6780}}
\affiliation{Dipartimento di Fisica, Sapienza Università di Roma, 00185 Roma, Italy}
\affiliation{Istituto Nazionale di Fisica Nucleare, Sezione di Roma, 00185 Roma, Italy}

\author{C.~M.~B.~Monteiro\orcid{0000-0002-1912-2804}}
\affiliation{LIBPhys, Department of Physics, University of Coimbra, 3004-516 Coimbra, Portugal}

\author{R.~A.~Nobrega\orcid{0000-0001-5199-308X}}
\affiliation{Universidade Federal de Juiz de Fora, Faculdade de Engenharia, 36036-900 Juiz de Fora, MG, Brazil}

\author{I.~F.~Pains\orcid{0009-0004-0851-6308}}
\affiliation{Universidade Federal de Juiz de Fora, Faculdade de Engenharia, 36036-900 Juiz de Fora, MG, Brazil}

\author{E.~Paoletti}
\affiliation{Istituto Nazionale di Fisica Nucleare, Laboratori Nazionali di Frascati, 00044 Frascati, Italy}

\author{F.~Petrucci\orcid{0000-0002-5278-2206}}
\affiliation{Dipartimento di Matematica e Fisica, Università Roma Tre, 00146 Roma, Italy}
\affiliation{Istituto Nazionale di Fisica Nucleare, Sezione di Roma Tre, 00146 Roma, Italy}

\author{S.~Piacentini\orcid{0000-0002-1256-7149}}
\affiliation{Gran Sasso Science Institute, 67100 L'Aquila, Italy}
\affiliation{Istituto Nazionale di Fisica Nucleare, Laboratori Nazionali del Gran Sasso, 67100 Assergi, Italy}

\author{D.~Pierluigi}
\affiliation{Istituto Nazionale di Fisica Nucleare, Laboratori Nazionali di Frascati, 00044 Frascati, Italy}

\author{D.~Pinci\orcid{0000-0002-7224-9708}}
\affiliation{Istituto Nazionale di Fisica Nucleare, Sezione di Roma, 00185 Roma, Italy}

\author{A.~A.~Prajapati\orcid{0000-0002-4620-440X}}
\altaffiliation{Present address: University of L’Aquila, Edificio Renato Ricamo, via Vetoio, Coppito, 67100 L’Aquila, Italy}
\affiliation{Gran Sasso Science Institute, 67100 L'Aquila, Italy}
\affiliation{Istituto Nazionale di Fisica Nucleare, Laboratori Nazionali del Gran Sasso, 67100 Assergi, Italy}

\author{F.~Renga\orcid{0000-0001-8129-8504}}
\affiliation{Istituto Nazionale di Fisica Nucleare, Sezione di Roma, 00185 Roma, Italy}

\author{A.~Russo}
\affiliation{Istituto Nazionale di Fisica Nucleare, Laboratori Nazionali di Frascati, 00044 Frascati, Italy}

\author{G.~Saviano}
\affiliation{Istituto Nazionale di Fisica Nucleare, Laboratori Nazionali di Frascati, 00044 Frascati, Italy}
\affiliation{Dipartimento di Ingegneria Chimica, Materiali e Ambiente, Sapienza Università di Roma, 00185 Roma, Italy}

\author{P.~A.~O.~C.~Silva\orcid{0000-0002-1957-2274}}
\affiliation{LIBPhys, Department of Physics, University of Coimbra, 3004-516 Coimbra, Portugal}

\author{N.~J.~C.~Spooner}
\affiliation{Department of Physics and Astronomy, University of Sheffield, Sheffield S3~7RH, United Kingdom}

\author{R.~Tesauro\orcid{0009-0006-0722-5896}}
\affiliation{Istituto Nazionale di Fisica Nucleare, Laboratori Nazionali di Frascati, 00044 Frascati, Italy}

\author{S.~Tomassini\orcid{0000-0001-7290-2028}}
\affiliation{Istituto Nazionale di Fisica Nucleare, Laboratori Nazionali di Frascati, 00044 Frascati, Italy}

\author{S.~Torelli\orcid{0000-0003-3622-3524}}
\altaffiliation{Present address: Donostia International Physics Center, BERC Basque Excellence Research Centre, Manuel Lardizabal 4, 20018 San Sebastián/Donostia, Spain}
\affiliation{Gran Sasso Science Institute, 67100 L'Aquila, Italy}
\affiliation{Istituto Nazionale di Fisica Nucleare, Laboratori Nazionali del Gran Sasso, 67100 Assergi, Italy}

\author{D.~Tozzi\orcid{0009-0001-9206-7354}}
\affiliation{Dipartimento di Fisica, Sapienza Università di Roma, 00185 Roma, Italy}
\affiliation{Istituto Nazionale di Fisica Nucleare, Sezione di Roma, 00185 Roma, Italy}

\date{\today}

\begin{abstract}
We report the first observation of Negative Ion Drift (NID) at surface pressure of $900 \pm 7$ mbar at Laboratori Nazionali del Gran Sasso in a He:CF$_4$:SF$_6$ mixture using an optically read out Time Projection Chamber (TPC) within the CYGNO/INITIUM project. We present the first PMT waveform analysis in the NID regime, interpreting the temporal light pattern through a model that combines track geometry and charge transport. The inferred drift velocities correspond to mobilities of  O(cm$^2$ V$^{-1}$ s$^{-1}$), consistent with negative ion transport. The observed linear scaling of the time extension mean with drift distance reveals the presence of a faster minority charge carrier population in addition to the dominant SF$_6^-$ species, drifting at a $\sim$25\% higher velocity under external inputs.
These results demonstrate multi-species negative ion drift operation at surface pressure in a He:CF$_4$:SF$_6$ mixture and open a concrete path toward large scale, low diffusion optical TPCs for rare event searches.

\end{abstract}

\maketitle

\section{\label{sec:intro}Introduction}

Charge diffusion during drift represents a well-known limitation of large gaseous Time Projection Chambers (TPCs), as it degrades spatial resolution and track reconstruction over extended volumes. While strong magnetic fields can suppress transverse diffusion \cite{Blum_rolandi}, their cost and complexity severely limit scalability. An alternative solution is provided by Negative Ion Drift (NID), where electrons produced by primary ionization are rapidly captured by an electronegative dopant and drift as negative ions  \cite{Martoff:2000wi,Ohnuki:2000ex,Phan2016TheNP}. Owing to their  larger mass, negative ions remain close to thermal equilibrium with the gas, yielding diffusion near the thermal limit without the need for magnetic fields.
Beyond diffusion suppression, several NID gas mixtures exhibit the simultaneous transport of multiple negative ion species with different mobilities \cite{Snowden-Ifft:2014taa,Phan2016TheNP}. The resulting separation in arrival times provides intrinsic timing information that can enable 3D fiducialization without  knowledge of the event time absolute. These feature make NID  attractive for rare event searches experiments requiring large target volumes and high quality tracking, as directional dark matter searches \cite{Battat:2016pap, Vahsen:2020pzb} and low energy solar neutrino spectroscopy \cite{Torelli:2024mof,Lisotti:2024fco}. 

Practical NID operation has been historically limited by suitable electronegative dopants (i.e. CS$_2$ \cite{Martoff:2000wi,Ohnuki:2000ex}, CH$_3$NO$_2$ \cite{Dion, Prieskorn:2014qga}), which raise toxicity and stability concerns. More recently, SF$_6$ has been demonstrated  capable to sustain NID operation, while providing a chemically stable and safe alternative \cite{Phan2016TheNP}, combined with high sensitivity to spin-dependent dark matter interactions through its fluorine content. Although SF$_6$-based NID has been demonstrated at reduced pressure with charge readout \cite{Ikeda:2020pex,Baracchini:2017ysg,Amaro:2024hom}, no observation had yet been reported at surface pressure with an optical readout, and direct
evidence of multi-population transport in He- or CF$_4$-based mixtures is still missing.

In this Letter, we present the first optical TPC operation in Negative Ion Drift regime at Laboratori Nazionali del Gran Sasso (LNGS) surface pressure in a He:CF$_4$:SF$_6$ gas mixture, achieved within the CYGNO/INITIUM project \cite{Amaro:2022gub}. By combining scientific CMOS cameras imaging, a trigger strategy based on Gas Electron Multiplier (GEM) signals, and PMT waveforms, we demonstrate unambiguous signatures of negative ion transport at surface pressure. An original timing algorithm applied to NID PMT waveforms acquired at 650 mbar over varying drift distances and electric fields reveals behavior consistent with at least two negative-ion populations with distinct mobilities, separated at the $\sim$25\% level. These advances directly address the scalability and readout flexibility required for next generation rare events large volume TPCs.

\section{\label{sec:atmospheric}Evidence of NID operation at surface pressure with optical readout}

The measurements at the LNGS surface pressure ($900 \pm 7$ mbar) were performed with the MANGO detector developed within the CYGNO/INITIUM program \cite{Amaro:2022gub,initium}. MANGO is a $10\times10$~cm$^2$ readout area optical Time Projection Chamber described in detail in Refs.~\cite{Baracchini:2020dib,Amaro:2024sms}. The detector was operated in continuous gas flow using the standard He:CF$_4$ 60:40 mixture for electron drift (ED) and the He:CF$_4$:SF$_6$ 59:39.4:1.6 mixture for negative ion drift (NID). The latter choice was motivated by previous charge readout studies demonstrating NID operation in the same gas mixture proportion at reduced pressure \cite{Baracchini:2017ysg}.

Ionization tracks were generated by a $^{241}$Am source emitting 5.485~MeV $\alpha$ particles, placed outside the active volume between the field cage rings at the center of the 5 cm drift gap. In ED operation, the three GEMs were biased at 310~V each, with a transfer field of 2.5~kV/cm and a drift field of 700~V/cm. In NID operation, the GEM voltages were set to 550/545/540~V (GEM1/GEM2/GEM3), while maintaining the same drift and transfer fields, in order to equalize the light yield between the two gas mixtures while ensuring stable operation.

\begin{figure}[!t]
    \includegraphics[width=0.65\linewidth]{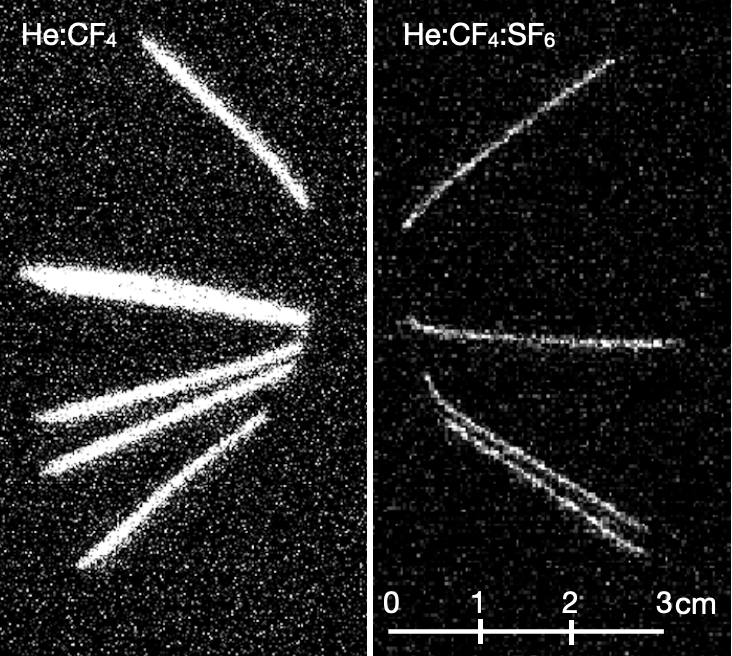}
    \caption{Example of raw scientific CMOS camera images acquired with MANGO exposed to a $^{241}$Am source and operated with He:CF$_4$ 60:40 (left) and He:CF$_4$:SF$_6$ 59:39.4:1.6 (right) at LNGS surface pressure. Images are displayed at the same pixel scale with no rebinning, with a scale bar shown for reference.}
    \label{fig:alpha}
\end{figure}

The sCMOS images were acquired in trigger-less mode with a 0.5~s exposure time. Representative examples for ED and NID operation are shown in Fig.~\ref{fig:alpha}, with identical pixel size. A striking difference in track morphology is observed between the two gas mixtures. Since the amplification settings were chosen to provide a comparable light yield per track, this difference cannot be simply attributed to gain or light production effects and instead reflects the distinct charge transport regime, providing a first qualitative indication of negative ion drift at surface pressure.

\begin{figure}[!t]
    \includegraphics[width=0.98\linewidth]{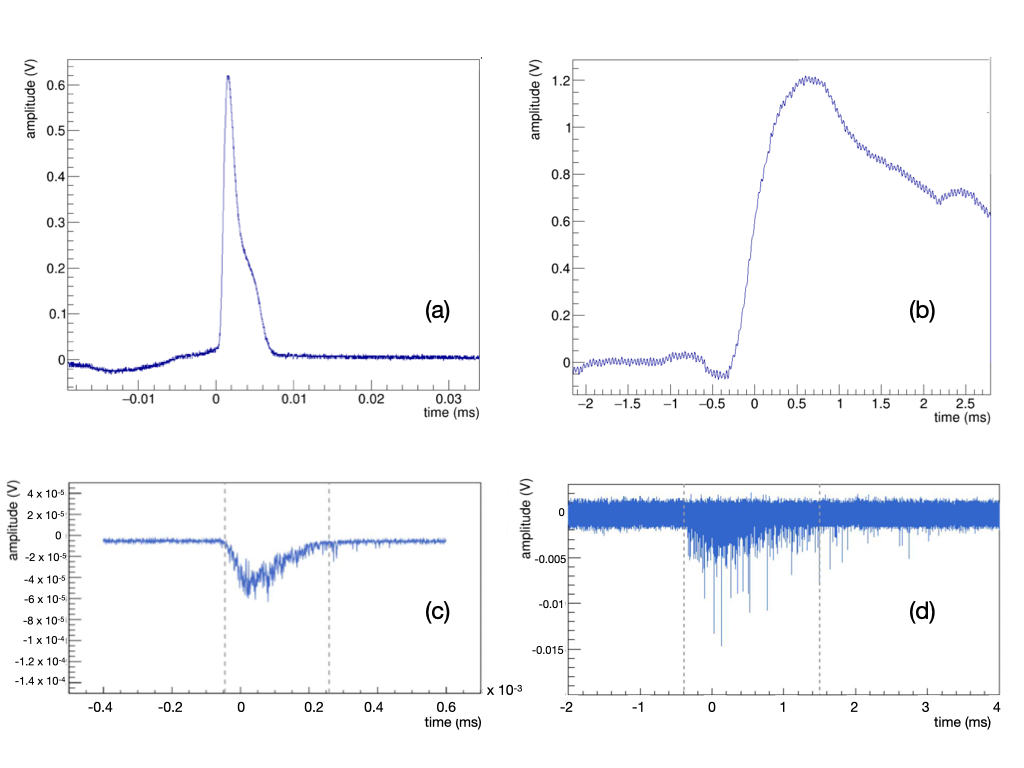}
    \caption{Charge signal and PMT response from an alpha track for electron drift (ED, left) and negative ion drift (NID, right) operation at LNGS surface pressure. Top: GEM3 preamplifier output, showing the fast and compact charge arrival in ED (a) and the millisecond scale arrival in NID (b). Bottom: Corresponding ED (c) and NID (d) PMT waveforms, highlighting the compact electron drift signal versus the sparse, extended time structure characteristic of negative ion transport. The dashed lines represent the end and start of the signal, as identified by the algorithm illustrated in Sec. \ref{subsec:pmt650}.}
    \label{fig:trigger_signal}
\end{figure}

The PMT signals and the charge collected on the bottom of the last GEM electrode (GEM3) were acquired using a highly performing Teledyne LeCroy oscilloscope (WR 9404-MS model), the latter after pre-amplification with a CAEN A422A (50 ns rise time, 300 $\mu$s decay time). In ED operation, the compact PMT signals allow standard triggering on the PMT waveform negative edge. In contrast, NID operation produces sparse, single-photoelectron-like PMT signals extending over several milliseconds, rendering PMT-based triggering ineffective. An robust acquisition strategy for NID events was therefore implemented by triggering on the GEM3 charge signal, which integrates the arrival of drifting anions and provides a clean and reliable event trigger. In ED operation, this signal exhibits a fast rise time of order hundreds of nanoseconds, reflecting the nearly simultaneous arrival of electrons, as shown in Fig. \ref{fig:trigger_signal} (a). In NID operation, the same signal develops over several milliseconds as evident in Fig. \ref{fig:trigger_signal} (b), directly mirroring the slow arrival time distribution of drifting anions and providing a clean and reliable trigger for NID events.

Figure~\ref{fig:trigger_signal} bottom panel shows typical PMT waveforms acquired in ED (c) and NID (d) operation. In the ED case, the signal is confined to a time window of a few hundred ns, consistent with typical electron drift velocities. In contrast, the NID waveform comprises a sequence of small peaks extending over several milliseconds, reflecting the much lower ion drift velocities. Although the short 5 cm drift distance in this configuration prevents a quantitative mobility extraction, the observed temporal broadening of the PMT signal and the associated slower rise time of the GEM3 output provide clear evidence that the charge carriers are negative ions rather than electrons.
Taken together, the images and waveforms presented in Fig. \ref{fig:alpha} and Fig. \ref{fig:trigger_signal} establish the first NID operation with an optical TPC at surface pressure and motivate the quantitative study of PMT timing observables presented in the following section.

\section{NID operation at 650 mbar}\label{sec:driftmobkeg}

In order to study NID PMT waveforms dependences on drift distance and electric field, measurements were performed at reduced pressure using an extended drift configuration. The same amplification stage used at surface pressure was equipped with a 15 cm field cage and installed inside a 150 L stainless steel vessel previously used for charge readout measurements (since it would not fit in the standard MANGO gas vessel) \cite{Baracchini:2017ysg}. In this configuration, referred to as MANGOk, the optical sensors were placed outside the vessel and coupled to the GEM plane through a quartz window with 90\% transparency. The detector was operated at a pressure of $(650 \pm 1)$ mbar with the same gas compositions used at surface pressure for both electron drift (ED) and negative ion drift (NID). The reduced pressure allowed compensation for the smaller optical acceptance of the external sensors imposed by the vessel design (about 1/3 with respect to the surface pressure setup) by increasing the effective gain while preserving stable operation. 

Ionization tracks were produced with the same $^{241}$Am source, further collimated by a slit of $\sim$ 0.3 cm vertical height to restrict the angular spread and enhance populations of alpha tracks nearly perpendicular to the drift direction. By assuming spherical coordinates $(\phi,\theta)$ with polar axis in the GEM plane, such collimation selects tracks uniformly distributed in $\phi\in[-\phi_{\rm max},\phi_{\rm max}]$ and in $\cos\theta\in[-1,1]$, forming a thin vertical fan rather than a forward cone. NID data were collected at five average drift distances ($Z_{\rm av}=2.5$, 3.5, 4.5, 6.5 and 9.5~cm), and four drift fields ($E_d = 300$, 350, 400 and 600~V/cm), with the GEMs powered to 535/530/525 V. A reference dataset in ED operation was acquired at $Z_{\rm av}=3.5$ cm and $E_d=300$ V/cm with each GEM biased at 305 V. The PMT acquisition and triggering scheme followed the same approach described in Sec.~\ref{sec:atmospheric}.

\subsection{PMT waveform analysis}\label{subsec:pmt650}

For each event, the PMT waveform time extension $\Delta T$ is defined as the time interval between the first and last portions of the waveform identified as signal. The extraction procedure reflects the markedly different temporal structure of the PMT signals observed in electron drift  and negative ion drift operation (Fig. \ref{fig:trigger_signal}).

In the ED regime, the PMT waveform is compact and confined to a time window of a few hundred nanoseconds. In this case, the electronic noise RMS is evaluated from the first 100 recorded samples, which is not expected to contain any signal, and $\Delta T$ is defined by identifying the first (last) bin at which the absolute waveform amplitude exceeds (falls below) 3 times the noise RMS to ensure high efficiency while suppressing noise-induced fluctuations.

In the NID regime, the PMT signal consists of sparse, single photoelectron-like peaks distributed over millisecond time scales. A dedicated algorithm was therefore developed to robustly identify the signal time extension in the presence of electronic noise and discrete nature of NID signals. The waveform noise is estimated from an initial signal-free region chosen to be significantly longer than in the ED case to account for the extended temporal structure of NID signals. A rebinned time profile over the acquisition window is built using only waveform peaks exceeding a fixed multiple of the RMS, as detailed in the Supplemental Material, in order to suppress contributions from baseline electronic noise while preserving the global
temporal structure of the signal. On the rebinned profile, the signal time extension $\Delta T$ is defined through a threshold-based identification of its start and end, requiring two consecutive bins to exceed (or fall below) the threshold in order to suppress sensitivity to statistical fluctuations.

The stability of the extracted $\Delta T$ values against the analysis choices is evaluated by varying the rebinning and threshold parameters over a conservative range, resulting in an overall systematic uncertainty of 20\% on $\Delta T$, which is propagated to the subsequent analysis (see Sec.~\ref{sec:mob}). Further details of the waveform analysis procedure, including an illustrative example, are provided in the Supplemental Material.

\subsection{Physical interpretation of PMT waveforms}
\label{sec:interp}
The PMT waveform encodes the arrival time distribution of the primary ionization charge carriers at the last amplification stage (i.e. GEM3 in our setup), which is converted into scintillation light during avalanche multiplication and detected by the PMT. Since the light emission process occurs on time scales negligible compared to the drift times and is localized to the avalanche region, the photons collected by the PMT directly reflect the temporal structure of the primary charge arrival. As a result, an alpha particle releasing an ionization pattern with range $R$ in the active gas volume will produce a PMT waveform signal spanning a time extension $\Delta T$ of about
 \begin{equation}\label{eq:dt_0}
\Delta T(\phi) \simeq \frac{R |\sin\phi|}{v_d} = \frac{|\Delta Z (\phi)|}{v_d}, 
\end{equation}
assuming the system of coordinate introduced in Sec. \ref{sec:atmospheric} and charge carriers drifting with velocity $v_d=\mu E_d$, where $\mu$ is the charges mobility. Since the source emission is isotropic, a time-projection measurement of an extended alpha track introduces a non-trivial Jacobian due to the transformation $\phi\!\to\!\Delta T$. This naturally yields large probability densities at small $\Delta T$ and a monotonically decreasing tail vanishing at the geometric endpoint $\Delta T_{\max}^{\rm geom}$ set by the collimation cone. 

Beyond these phenomenological considerations, the observed time extension reflects the collective behavior of many charge carriers produced along the track, whose individual drift and amplification times are governed by stochastic transport processes. Since $\Delta T$ measures the temporal separation between the earliest and latest carrier arrival, it results governed by the tails of the underlying arrival time distributions through the extreme-value statistics (EVT)\cite{Majumdar:2020} rather than by their bulk properties.

If all charge carriers drift with the same velocity, as in the electron drift regime, increasing the drift distance rigidly shifts the entire arrival time distribution to later times, while its width grows through longitudinal diffusion. In this case, the separation between the earliest and latest arrivals arises from geometry and diffusion only, exhibiting the characteristic $\sqrt{Z_{\rm av}}$ dependence. In contrast, when multiple charge populations with different mobilities coexist, the earliest and latest arrivals are statistically dominated by the fastest and slowest carriers, respectively. Their temporal separation therefore accumulates not only through stochastic diffusion but also deterministically with drift distance, due to the difference in drift velocities. This introduces an additional linear dependence of $\Delta T$ on $Z_{\rm av}$, superimposed on the geometric baseline. No mechanism based solely on diffusion or amplification smearing can generate such a linear term. 
In this regime, the residual event-to-event fluctuations around the characteristic $\Delta T$ are primarily driven by longitudinal diffusion and detector response, both of which are smooth stochastic processes. As a result, the distribution of the measured $\Delta T$ in the NID case are empirically well described by a Gaussian function. A quantitative derivation of the single and multi-species behavior for the experimental configuration presented here, including the full geometric treatment and extreme-value formalism, is provided in the Supplemental Material.

\subsection{NID mobilities estimate}
\label{sec:mob}
The NID time extension $\Delta T$ is evaluated event by event as described in Sec.~\ref{subsec:pmt650}. For each drift distance and drift field configuration, the corresponding $\Delta T$ distributions results well described by a single Gaussian over the full range explored, with an average displaying a clear and systematic linear trend on $Z_{\rm av}$. We therefore define the average measured time extension $\langle \Delta T \rangle(Z_{\rm av},E_d)_m$ as the mean returned by a Gaussian fit to each distribution.

As illustrated in Sec.~\ref{sec:interp}, in the NID regime the average of the $\Delta T$ distributions is controlled by the temporal separation between extreme charge arrivals and therefore expected to scale inversely with the drift velocity. Using $v_d=\mu E_d$, this motivates the use of the rescaled observable $E_d\langle \Delta T \rangle_m$, which removes the trivial leading field dependence and allows the data acquired at different drift fields to be combined, while preserving a common systematic uncertainty associated with the absolute drift distance for each $Z_{\rm av}$.

For each $(Z_{\rm av},E_d)$ configuration, the uncertainty is obtained by combining the statistical error on the mean returned by the Gaussian fit to the $\Delta T$ distribution with the uncertainty on the applied drift field $E_d$, conservatively taken as $\sigma_E=5$~V/cm and treated as a per-setting (point-to-point) uncertainty across the drift field scan. For each $Z_{\rm av}$, the four drift field measurements are combined through a weighted average and the assigned error is obtained with standard uncertainty propagation. The uncertainty on $Z_{\rm av}$ is conservatively set to $0.3$~cm, dominated by the finite
extent of the source relative to the collimation slit aperture (see Sec.\ref{sec:driftmobkeg}), with only a subdominant
contribution from positioning.

\begin{figure}[!t] 
 \includegraphics[width=0.8\linewidth]{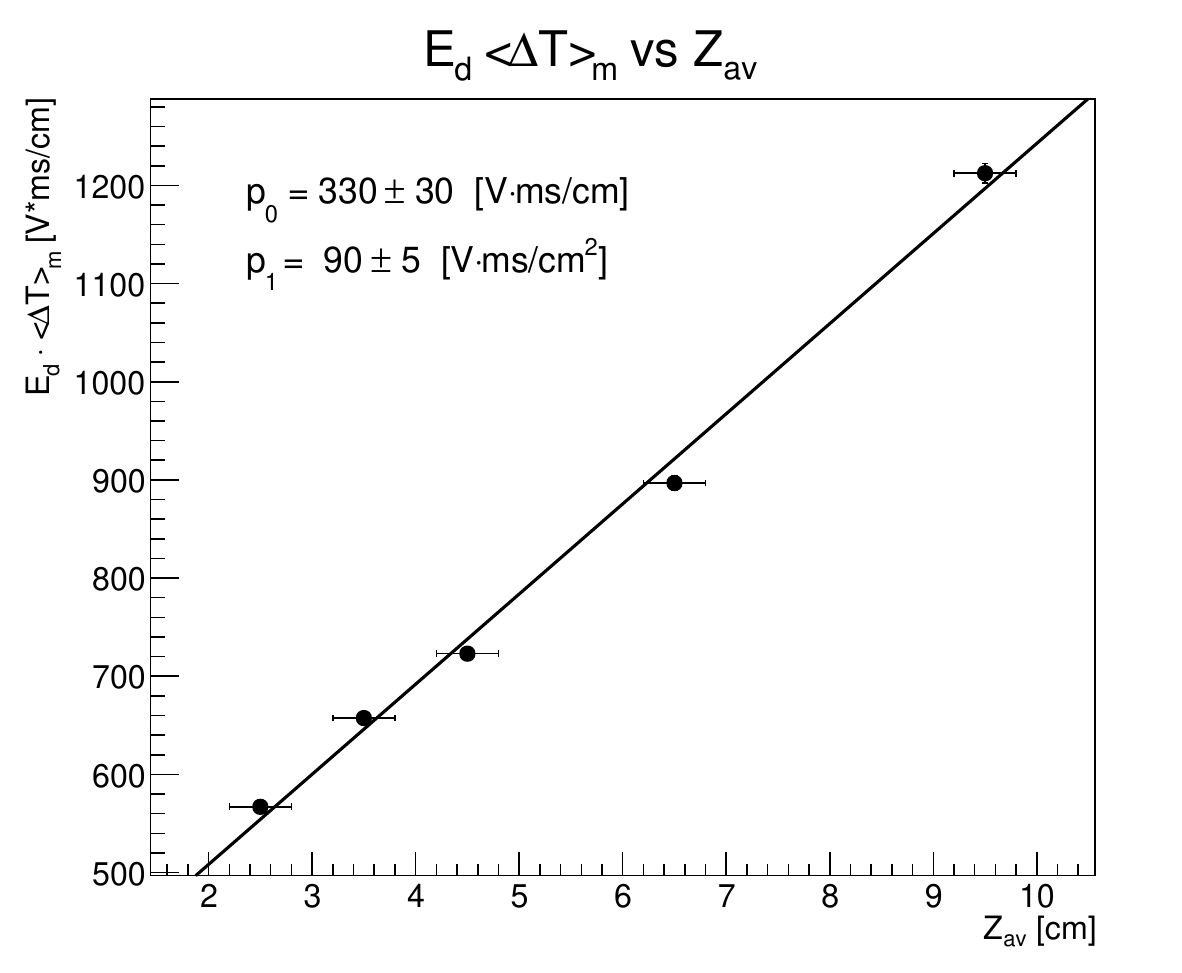}
	\caption{Averaged $E_d \langle \Delta T \rangle$ as a function of the drift distance $Z_{\rm av}$ with a linear fit superimposed ($\chi^2/\mathrm{ndf}=1.69/3$).}
	\label{fig:nid_fit}
\end{figure}

Figure~\ref{fig:nid_fit} shows the rescaled averaged observable $E_d\langle \Delta T\rangle_m$ as a function of the drift distance $Z_{\rm av}$. The data collapse onto a common trend and are well described by the linear relation
\begin{equation}
E_d \langle \Delta T \rangle_m = p_0 + p_1 Z_{\rm av},
\label{eq:scaling_fit}
\end{equation}
where a least-squares fit including both vertical and horizontal uncertainties returns
\begin{align}\label{eq:nid_res}
p_0 &= 330 \pm 30_{\rm stat} \pm 60_{\rm syst} \,\,[{\rm V \cdot ms/cm}]\\ \nonumber p_1 &= \, \,\,90 \pm\,\, 5_{\rm stat}\, \pm 20_{\rm syst} \,\,[{\rm V \cdot ms/cm^2}].
\end{align}
with a $\chi^2/\mathrm{ndf}=1.69/3$, indicating good agreement with a linear scaling. Here we assigned a conservative 20$\%$ systematic uncertainty on $p_0$ and $p_1$ evaluation, to account for the stability of the
$\Delta T$ extraction procedure (e.g. binning and threshold choices discussed in Sec.~\ref{subsec:pmt650}). This contribution is treated as an overall scale uncertainty on $E_d\langle\Delta T\rangle_m$, and is therefore propagated multiplicatively to the fitted parameters. Despite this approach does not represent the most rigorous method to estimate such systematics, it provides a conservative and physically motivated estimate of the dominant contribution, ensuring that the quoted mobility
values reflect the actual level of experimental constraint.

As discussed in Sec.~\ref{sec:interp}, the emergence of a linear term in $Z_{\rm av}$ is a compelling signature of multi-population transport in the NID regime and provides a direct handle on mobility differences. Within the transport framework illustrated in details in the Supplementary Material, the fit parameters can be interpreted as:
\begin{align}
p_0 &= \left(\frac{1}{\mu_s}+\frac{1}{\mu_f}\right)\!\left(R|\sin\phi_{\rm eff}|+\sigma_{0L}\sqrt{2\ln N_t^{\rm eff}}\right), \nonumber\\
p_1 &= \left(\frac{1}{\mu_s}-\frac{1}{\mu_f}\right),
\end{align}
where $\mu_s$ and $\mu_f$ denote the mobilities of the slowest and fastest drifting charge populations,  $\sigma_{0 L}$ their diffusion at the amplification plane, $|\sin\phi_{\rm eff}|$ the effective angular weighting and N$^{\rm eff}_t$ the number of primary charges contributing to the arrival track smearing at the track borders, weighted by the convolution with the geometrical factor.

Although the geometric factor $R|\sin\phi_{\rm eff}|$ cannot be measured directly, its magnitude can be constrained from the ED data and the collimation geometry of the setup, yielding a characteristic scale of order O(1)~cm. Using this estimate together with the expected values of $\sigma_{0L}$ and $N_t^{\rm eff}$, the fitted values of $p_0$ correspond to mobilities of order O(1)~cm$^2$\,V$^{-1}$\,s$^{-1}$. These values are nicely consistent with typical negative-ion transport properties and with the reduced SF$_6^-$ mobility reported in Ref.~\cite{Baracchini:2017ysg}. 

Despite the number of unknown quantities (i.e. R$|\sin\phi_{\rm eff}|$, $N_t^{\rm eff}$, $\sigma_{0L}$, $\sigma_{L}$) do not allow to extract $\mu_s$ and $\mu_f$ independently, under the assumption that the slow component corresponds to the dominant SF$_6^-$ species and using previously measured transport properties as an external reference \cite{Baracchini:2017ysg}, we estimate the reduced mobility of the fast charge carriers observed in the data from the fitted p$_1$ as:
\begin{equation}
    \mu^{f}_0 = 3.0 \pm 0.1_{\rm  stat} \pm 0.2_{\rm  syst} \text{ cm}^2 \text{ V}^{-1} \text{ s}^{-1},
\end{equation}
implying a mobility about 25$\%$ higher than the slower SF$_6^-$ component. The quoted uncertainty on $\mu_0^f$ is obtained by standard propagation of the statistical and systematic uncertainties on $p_1$, together with the uncertainty on the external input $\mu_s$ from Ref.~\cite{Baracchini:2017ysg}. Given the measured linear scaling and the independently constrained value of $\mu_s$, the extracted $\mu_0^f$ constitutes a quantitatively robust, though indirect, determination of the fast carriers mobility within the present framework. 

\section{Conclusions}

We have reported the first observation of Negative Ion Drift at surface pressure in a He:CF$_4$:SF$_6$ gas mixture using an optical Time Projection Chamber, and the first demonstration that NID can be directly detected and characterized through photomultiplier waveforms. By combining sCMOS imaging, GEM-based triggering, and a dedicated PMT waveform analysis, we demonstrated stable NID operation at 900~mbar and performed a quantitative study at 650~mbar over different drift distances and multiple electric fields.

By developing a dedicated original analysis for the sparse, millisecondscale PMT signals characteristic of negative ion drift, we extracted the time-extension observable $\langle \Delta T \rangle_m$ encoding information on charge transport and track geometry. Measurements performed at 650 mbar over multiple drift distances and electric fields reveal a systematic linear dependence of the rescaled observable  $E_{d}\langle \Delta T \rangle_m$ on drift distance. To provide a quantitative interpretation of this behaviour, we developed a general physical framework in which the PMT waveform time extension is expressed as the convolution of a macroscopic geometric projection of extended tracks with a microscopic transport kernel governed by longitudinal diffusion and extreme-value statistics. This approach, fully derived in the Supplemental Material, offers a novel unified description of PMT waveform time extensions applicable to both electron drift and negative ion drift regimes. Within this framework, we showed that the linear increase of the $\Delta T$ distribution mean with $Z_{\rm av}$ cannot be explained by geometry or diffusion alone and constitutes direct experimental evidence for the coexistence of multiple negative ion species with different mobilities. Under the assumption that the slowest component observed coincides with the dominant SF$_6^{-}$ charge carrier and employing external inputs, we
estimate the presence of a faster minority carrier drifting at approximately 25\% higher velocity.

This work demonstrates for the first time that multi-species negative ion drift at surface pressure can be directly observed and quantitatively characterized through PMT waveform diagnostics.  
These results establish optical TPCs as a viable and scalable platform for next-generation directional detectors, where the combined use of imaging and time-domain information enables enhanced fiducialization, background rejection, and sensitivity in rare event searches.

\begin{acknowledgments}
This project has received fundings under the European Union’s Horizon 2020 research and innovation programme from the European Research Council (ERC) grant agreement No 818744. We want to thank General Services and Mechanical Workshops of Laboratori Nazionali di Frascati (LNF) and Laboratori Nazionali del Gran Sasso (LNGS) for their precious work and L. Leonzi (LNGS) for technical support.
\end{acknowledgments}

\bibliography{apssamp}

%

\clearpage

\onecolumngrid
\setcounter{figure}{0}
\setcounter{table}{0}
\setcounter{equation}{0}
\setcounter{section}{0}
\setcounter{subsection}{0}

\renewcommand{\thefigure}{S\arabic{figure}}
\renewcommand{\thetable}{S\arabic{table}}
\renewcommand{\theequation}{S\arabic{equation}}
\renewcommand{\bibnumfmt}[1]{[S#1]}
\renewcommand{\citenumfont}[1]{S#1}
\begin{center}
{\Large\bf Supplemental Material for:}\\[0.3cm]
{\Large\bf First Optical Observation of Negative Ion Drift at Surface Pressure}\\[0.6cm]
\end{center}


F. D. Amaro, R. Antonietti, E. Baracchini, L. Benussi, C. Capoccia, M. Caponero, L. G. M. de Carvalho,
G. Cavoto, I. A. Costa, A. Croce, M. D’Astolfo, G. D’Imperio, G. Dho, F. Di Giambattista, E. Di Marco,
J. M. F. dos Santos, D. Fiorina, F. Iacoangeli, Z. Islam, H. P. Lima Jr., G. Maccarrone, R. D. P. Mano,
D. J. G. Marques, G. Mazzitelli, P. Meloni, A. Messina, C. M. B. Monteiro, R. A. Nobrega, I. F. Pains,
E. Paoletti, F. Petrucci, S. Piacentini, D. Pierluigi, D. Pinci, A. A. Prajapati, F. Renga, A. Russo,
G. Saviano, P. A. O. C. Silva, N. J. C. Spooner, R. Tesauro, S. Tomassini, S. Torelli, and D. Tozzi

\section*{Negative ion drift PMT waveform analysis}
This section presents the original analysis procedure developed to evaluate
the time extension of the PMT waveforms acquired  in NID operation with the MANGO detector setup described in Sec. II and Sec. III of the main paper. As illustrated, the fast response of the PMT, combined with the slow drift velocity of NID anions, manifests in a very sparse signal made of small, single photoelectron-like peaks. The large time span of these signals, combined with the small width of the individual peaks, requires recording at a high sampling rate on the order
of GS/s with the oscilloscope. However, this high sampling rate introduces additional
electronic noise, which can resemble the already low-amplitude signal peaks. To address
these challenges, an original algorithm was developed to manage both the electronic
noise introduced by the oscilloscope and the irregular, discrete nature of the NID signals. The purpose of this procedure is not to reconstruct individual charge carrier arrival times (or to perform single cluster counting), but rather to define a robust, global time extension observable, $\Delta T$, that can be consistently extracted event-by-event from sparse NID waveforms and used for transport and geometry studies. While the numerical thresholds and binning choices reported below are optimized for the present dataset (sampling rate, noise level, and typical NID occupancy), the procedure is intended to be generally applicable to NID configurations in which the PMT response is fast compared to the millisecond-scale drift and the waveform consists of sparse single-photoelectron–like peaks; only the tuning parameters (RMS threshold, rebinning, and consecutive-bin criterion) need to be adapted to the specific acquisition conditions.

\begin{figure}[!t] 
   \includegraphics[width=0.95\linewidth]{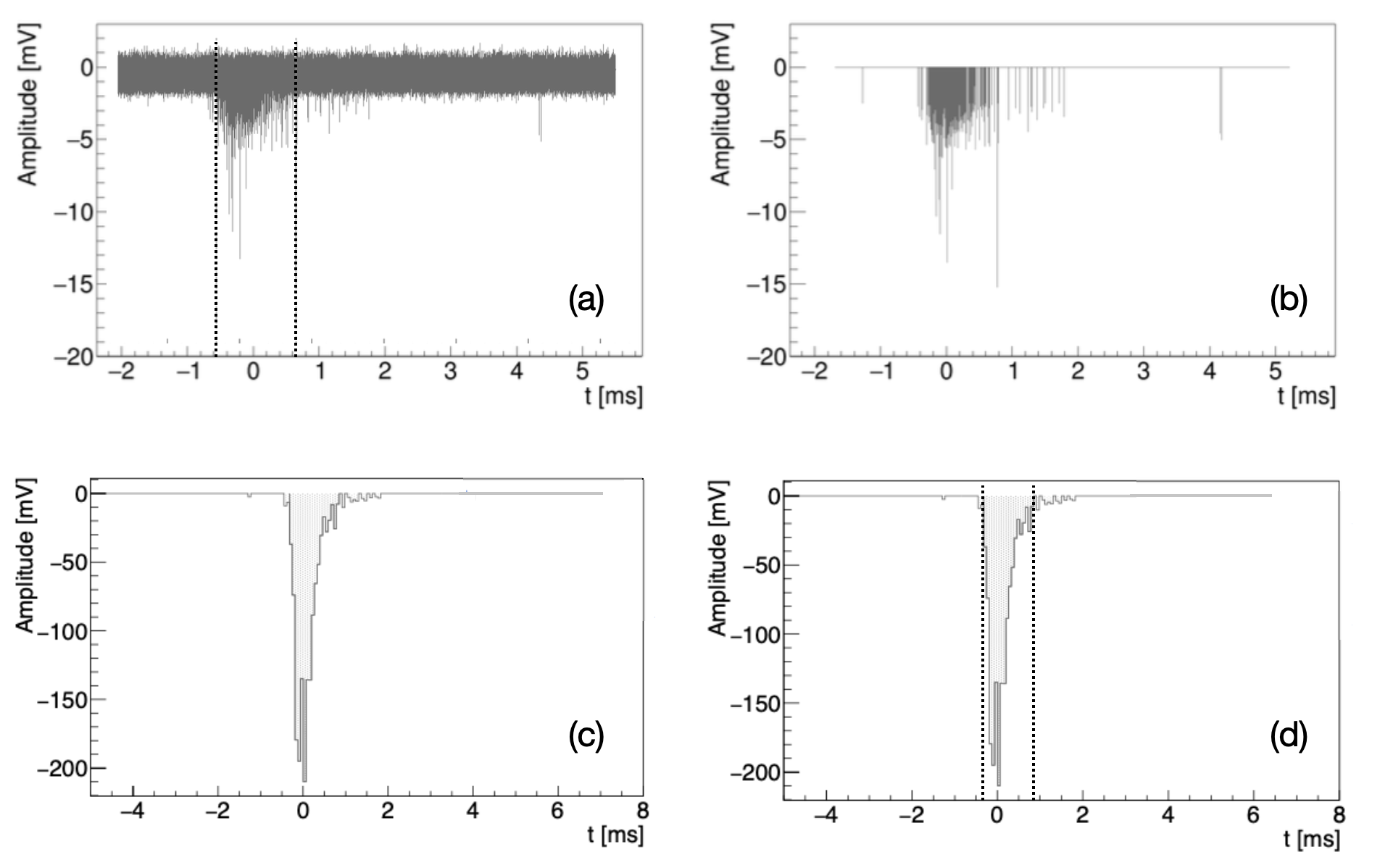}
	\caption{Example of NID PMT waveform analysis: (a) original NID waveform with \emph{start} and \emph{end} timestamps obtained with the illustrated procedure overlaid as black dashed lines, (b) NID waveform peaks above 6 times the waveform RMS, (c) rebinned histogram of top right waveform, (d) selected NID waveform $\Delta T$ interval between black dashed lines obtained defining the beginning (end) of the signal when two consecutive bins are above (below) 10 mV.}
	\label{fig:nid_ana}
\end{figure}

The analysis strategy proceeds as following:
\begin{enumerate}
    \item the first 500 recorded sample of the original raw waveform acquired with the Teledyne Oscilloscope (top left of Fig.\ref{fig:nid_ana}) which are expected to not contain any signal, are used to evaluate the RMS of the electronic noise. A region five times larger than in the ED case (see main text) is chosen to account for the extended temporal structure of the NID signals;

    \item from the original raw waveform, only peaks above 6 times the RMS, equivalent to $\sim$ 2.5 mV, are retained, where all the other points are set to zero, as shown in top right panel of Fig.\ref{fig:nid_ana}. This step results necessary because, due to the high sampling rate,  most of the points in the waveform correspond to the baseline electronic noise. Without this thresholding, a simple rebinning of the entire waveform would just blur the already few visible signal peaks. Since a NID waveform comprises $\mathcal{O}(10^{7})$ samples over $\mathcal{O}(10)$~ms, a $5\times$RMS threshold would still yield $\mathcal{O}(1)$ false positives per event  under a Gaussian-noise assumption. A $6\times$RMS criterion was therefore adopted, also accounting for possible  non-Gaussian tails in the PMT/electronics response.;

    \item the surviving peaks are used to build a rebinned time profile with 200 bins over a 10 ms extension (bottom left of Fig.\ref{fig:nid_ana}), from which the beginning (end) of the signal is defined when two consecutive bins of the rebinned time profile exceed (fall below) 10 mV (bottom right of Fig.\ref{fig:nid_ana} highlighted in red). The two-consecutive-bin threshold criterion is imposed in order to suppress sensitivity to statistical noise fluctuations.
    
\end{enumerate}

Threshold and rebinning parameters were chosen to minimize false positives in empty waveforms while ensuring sufficient occupancy per bin for typical NID signals. The start and end timestamps obtained with the above procedure are overlaid as gray dashed lines on the original raw waveform in top left of Fig.\ref{fig:nid_ana}. The algorithm was tested on a dedicated sample of O(100) empty NID waveforms, for which it returned a measured time extension consistent with zero, demonstrating its robustness against false positives.

The systematic uncertainties associated to this strategy are evaluated by repeating the analysis varying the number of bins of the rebinned histogram from 150 to 250 and the threshold on the two consecutive bins from 5 mV to 15 mV. Since the algorithm uncertainties may not be propagated analytically, by applying a confidence interval of 68\% on the distribution of the average of the NID $\Delta T$ time extensions obtained with such modified prescriptions, a systematic uncertainty of 20$\%$ is evaluated on the estimate of this quantity. This systematic is linearly propagated to the $p_0$ and $p_1$ values obtained from a fit to $E_d \langle\Delta T \rangle_m$ as a conservative uncertainty from the PMT waveform analysis procedure (see main text).

\begin{figure}[!t] 
   \includegraphics[width=0.5\linewidth]{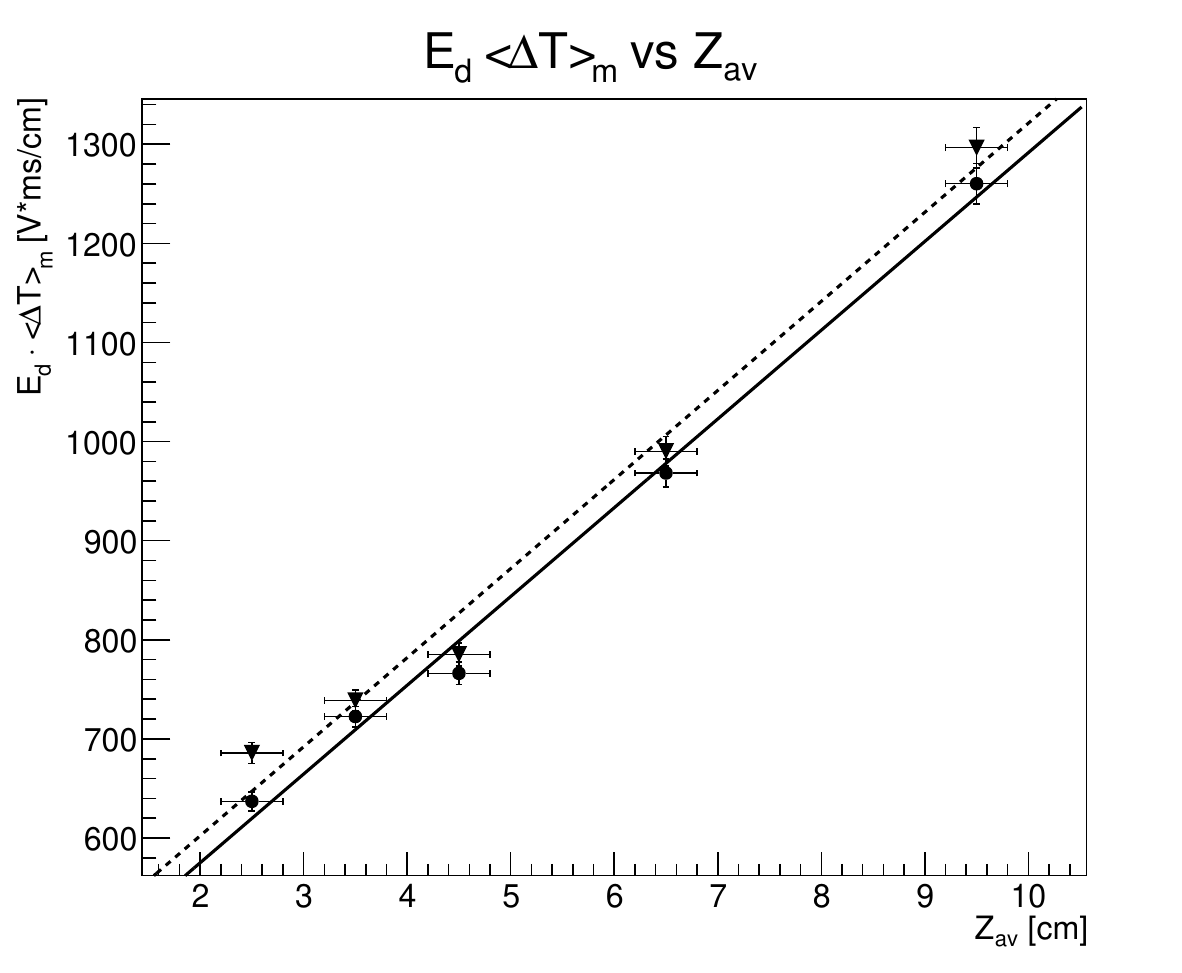}
	\caption{Rescaled observable $E_d\langle \Delta T \rangle$ 
as a function of the average drift distance $Z_{\rm av}$ 
at $E_d = 400$ V/cm for original data (black circle) and 
after injection of 10\% randomly distributed fake peaks 
(downward triangles). Linear fits are superimposed, with full line for original data and dashed line for data with random fake peaks injection. 
The fitted slope parameters are statistically compatible between them
within uncertainties and reported explicitly in the text.}
	\label{fig:nid_injected}
\end{figure}

\subsection*{Robustness against fake-peak injection}

To test whether the observed linear dependence of 
$E_d\langle \Delta T \rangle$ on $Z_{\rm av}$ could be 
artificially induced by sparse peak handling in the waveform 
analysis, we performed a dedicated fake-peak injection study. Starting from the noise-cleaned PMT waveforms (after applying 
the $6\times$ RMS threshold described above), we generate 
modified waveforms by adding an additional 10\% of peaks 
randomly distributed over the full acquisition time window 
of each event. The injected peaks are added to a copy of the 
original waveform, preserving the original signal structure. 
The complete $\Delta T$ extraction procedure (including 
rebinning, two-consecutive-bin thresholding, construction 
of the $\Delta T$ distributions, and linear fitting of 
$E_d\langle \Delta T \rangle$ versus $Z_{\rm av}$) is 
then repeated identically for all drift distances at fixed 
drift field ($E_d = 400$ V/cm). Figure~\ref{fig:nid_injected} shows the resulting 
$E_d\langle \Delta T \rangle$ versus $Z_{\rm av}$ 
for the original dataset (black points) and for the 
fake-peak injected sample (red points), together with 
their respective linear fits. For the original data we obtain:
\[
p_0 = 400 \pm 32~\text{V ms/cm}, \qquad
p_1 = 90 \pm 6~\text{V ms/cm}^2,
\]

while for the injected sample we obtain:
\[
p_0 = 420 \pm 32~\text{V ms/cm}, \qquad
p_1 = 91 \pm 6 ~\text{V ms/cm}^2.
\]

The fitted parameters obtained from the injected sample 
are found to be statistically compatible with those extracted 
from the original data. This study demonstrates that the observed linear trend is not generated by the sparse-peak treatment of the 
waveform analysis algorithm.


\section*{Full derivation of PMT waveform time extension observable}
This section derives a compact analytic representation of the measured time-extension distribution $p_m(\Delta T)$ as the convolution of (i) a macroscopic geometric projection of the track, set by the source collimation, and (ii) a microscopic transport kernel describing stochastic broadening of the earliest and latest arrivals. 
Only the resulting scaling with drift distance $Z_{\rm av}$ used in the main text (i.e. $\sqrt{Z_{\rm av}}$ for single-species transport and a linear $Z_{\rm av}$ term for multi-species drift-velocity separation) is required for the quantitative interpretation; intermediate steps are included to provide physical intuition and to make explicit the underlying assumptions.

Here we adopt the same geometrical configuration and coordinate system described in the main text (Sec.~II). In particular, the isotropic $^{241}$Am emission is collimated by a vertical slit of about $\sim$ 0.3 cm vertical height, uniformly distributed in a narrow angular range $\phi\in[-\phi_{\rm max},\phi_{\rm max}]$ and in $\cos \theta$ between $[-1,1]$. 

\subsection*{PMT time extension $\Delta T$ distribution}

As elaborated in the main text, given an isotropic source emission, the measurement of the track spatial projection along the drift direction $\Delta Z(\phi)$ from the PMT time extension $\Delta T (\phi)$ yields to a probability distribution $p_{\rm g}(\Delta T)$ shaped by the Jacobian of the transformation $\phi\!\to\!\Delta T$:
\begin{equation}\label{eq:geom}
p_{\rm g}(\Delta T)= \frac{v_d}{\phi_{\rm max}} \arccos \Big ( \frac{v_d\Delta T \cos \phi_{\rm max}}{|\sin \phi_{\rm max}| \sqrt{R^2-(v_d \Delta T)^2}}\Big)
\end{equation}
defined on
\begin{equation}\label{eq:geom_def}
0\le  \Delta T \le \Delta T^{\rm geom}_{\max}= \frac{|\Delta Z_{\rm max}|}{v_d} =\frac{R}{v_d}|\sin\phi_{\max}|.
\end{equation}

Microscopically speaking, the ionisation pattern results actually composed by a large number of primary carriers, whose charge density can be modeled as a tridimensional Gaussian distribution ~[S1] around the emission axis of the alpha particle (the track centerline). The initial $\rho_0$ distribution diffuse after drifting for a time $t$ along $z$ as
\begin{equation}
\rho(\mathbf{r}, t | \phi) = \int \rho_0(\mathbf{r}', \phi) \cdot G(\mathbf{r} - \mathbf{r}' - v_d t \mathbf{\hat{z}}; \sigma_{\rm diff}(t)) \, d\mathbf{r}',
\end{equation}
where $G$ is a multivariate Gaussian, with $\mathbf{r} = (x, y, z)$ and the drift term $v_d t \mathbf{\hat{z}}$ accounts for the displacement $v_d t = Z$ of charge carriers along the $z$-drift direction ~[S1].
The transverse $\sigma_{\rm diff,T}$ and longitudinal $\sigma_{\rm diff,L}$ sigmas can be defined as:

\begin{equation}\label{eq:s_diff}
    \sigma_{\rm diff,T/L} = \sqrt{\sigma_{0 \rm T/L}^2 +  \sigma_{\rm T/L}^2Z }
\end{equation}
where $\sigma_{\rm T/L}$ are the charge carriers diffusion coefficients and $\sigma_{0 \rm T/L}$ represent the smearing suffered at the amplification stage. Since charges drift along the longitudinal coordinate $z$, the transverse diffusion components can be marginalized and the temporal probability density $p_t(t|\phi)$ of a single charge carrier arrival at GEM3 after having drifted for $Z$ can be expressed as:
\begin{equation}\label{eq:pt}
p_t(t|\phi) \approx G(t; m_{t}, \sigma_t) = G\left(t; \frac{Z}{v_d}, \frac{\sqrt{\sigma_{0L}^2 + \sigma_L^2 Z}}{v_d}\right),
\end{equation}
 where the amplification plane is assumed at $z=0$, the mean represents the projected arrival time, and the standard deviation accounts for the longitudinal diffusion smearing.

The measured $\Delta T$ distribution $p_{m}(\Delta T)$ results therefore from the convolution of the angular probability density $p_{\rm g}(\phi) = 1/(2 \phi_{\rm max})$ with the temporal probability distribution $p_{\Delta T}(\Delta T|\phi)$ of the difference $\Delta T = T_{l} - T_{e}$ between the latest T$_l$ and earlier T$_e$ arrival times of $N$ charge carriers produced along the track, each following $p_t(t|\phi)$ distribution of Eq. \ref{eq:pt}, as:

\begin{align}\label{eq:p_conv}
    p_{m}(\Delta T) &= \int_{-\phi_{\rm max}}^{\phi_{\rm max}} \frac{ p_{\Delta T}(\Delta T  |\phi)}{2 \phi_{\rm max}} d \phi.
\end{align}

Each individual charge carriers drifts independently in the linear transport regime ~[S1], so that, to leading order, their arrival times can be treated as statistically independent.
Since Eq. \ref{eq:pt} is Gaussian and a Geant4 simulation of our setup shows that alphas from $^{241}$Am deposit O(10$^5$) primary ionisation clusters per cm, the extreme-value statistics (EVT) theory ~[S2] can be applied to estimate the difference between the maximum and minimum value of a set of $N$ independent samples drawn from $p_t(t|\phi)$. Within EVT, the expectation value of the difference of arrival times of the last and first carrier is equal to the difference of their relative expectation value as $\mathbb{E}[\Delta T] = \mathbb{E}[T_l] - \mathbb{E}[T_e]$ with:

\begin{align}
    \mathbb{E}[T_l] \simeq m_{t}^l + \sigma_t^l\sqrt{2 \ln N_{l}} \quad \mathbb{E}[T_e] \simeq m_{t}^e - \sigma_t^e\sqrt{2 \ln N_{e}}
\end{align}

in the large-N asymptotic regime relevant for alpha tracks. Here $m_{t}^{i}$ and $\sigma_{t}^{i}$ are the mean and sigma of the $p_t(t|\phi)$ distribution of Eq. \ref{eq:pt} for the latest/earlier ($i=l/e$) cluster and  $N_{e/l}$ is the number of  charge carriers effectively contributing to the arrival times smearing at the track borders, which depends on the geometry of the event. Since the residual fluctuations of the extreme arrival times arise from the convolution of many small stochastic contributions (diffusion and amplification smearing), their combined effect can be approximated as Gaussian at fixed $\phi$, even though the exact extreme-value distribution is not strictly Gaussian, as:
\begin{align}
     p_{\Delta T}(\Delta T) \simeq G(\Delta T; m_{\rm eff}, \sigma_{\rm eff})
\end{align}
with
\begin{align}
    m_{\rm eff}  \equiv  \mathbb{E}[T_l] - \mathbb{E}[T_e] \quad   \sigma_{\rm eff}^2 \simeq {\rm Var}(T_l) + {\rm Var}(T_e)
\end{align}
This formulation remains valid in the presence of multiple charge carriers with a non-degenerate distribution of drift velocities, since the earliest and latest arrivals are governed by the maximum and minimum velocities present in the ensemble.
The use of extreme-value statistics here should be understood as an effective description of the scaling behavior of the earliest and latest arrival times. Finite-sample effects are absorbed into an effective number of contributors $N_{\rm eff}$ without modifying the functional dependence of the extreme-value separation on drift distance.


\subsection*{Explicit derivation for the present experimental geometry}
In the setup described in the main paper, alpha tracks have all origin in $Z_{\rm av}$ by construction and end at $Z_{\rm av} + \Delta Z(\phi)$. The earlier (later) arrival time $T_e$ ($T_l$) can therefore correspond to the start or end of the track depending on the sign of $\Delta Z(\phi)$ and the EVT statistics. It results natural hence to split the integral of Eq. (\ref{eq:p_conv}) into positive and negative $\phi$ integration regions as:

\begin{align}\label{eq:p_split}
     p_{m}(\Delta T,Z_{\rm av}) = \frac{1}{2 \phi_{\rm max}} & \Big ( \int_{-\phi_{\rm max}}^{0} G(\Delta T; m_{\rm eff}^-, \sigma_{\rm eff}) d \phi + \int_{0}^{\phi_{\rm max}} G(\Delta T; m_{\rm eff}^+, \sigma_{\rm eff}) d \phi \Big )
\end{align}

with

\begin{align}
    m_{\rm eff}^{\pm}(Z_{\rm av},\phi) =   \Big (\frac{1}{v_l} - \frac{1}{v_e} \Big) Z_{\rm av}+ \sigma_t^l\sqrt{2 \ln N_{l}} +\sigma_t^e\sqrt{2 \ln N_{e}}+ \frac{\Delta Z (\phi)}{v_{l/e}},  \quad \quad
    \sigma_{\rm eff}(Z_{\rm av}) =  \sqrt{(\sigma_t^{l})^2 + (\sigma_t^{e})^2} .
\end{align}

where v$_e$ and v$_l$ are the earliest and latest arrival charge carrier drift velocities. Since the $\sigma_t^{l/e}$ terms depend on $\sqrt{Z_{\rm av}}$ (see Eq.\ref{eq:s_diff}), in order to separate the various dependences on Z$_{\rm av}$, we define

\begin{equation}
    \tau_0(Z_{\rm av}) = \Big (\frac{1}{v_l} - \frac{1}{v_e} \Big) Z_{\rm av}, \quad m_{\rm ex}(Z_{\rm av}) = \sigma_t^l\sqrt{2 \ln N_{l}} +\sigma_t^e\sqrt{2 \ln N_{e}}, \quad m_0(Z_{\rm av}) = \tau_0(Z_{\rm av}) + m_{\rm ex}(Z_{\rm av}), 
\end{equation}

,

so that we can write

\begin{align}
    m_{\rm eff}^{\pm}(Z_{\rm av},\phi) =   m_0(Z_{\rm av})+ \frac{\Delta Z (\phi)}{v_{l/e}}.
\end{align}
By applying the change of variables $ \tau^-(\phi) = \Delta Z (\phi)/v_e$ and $ \tau^+(\phi) = \Delta Z (\phi)/v_l$ in the negative and positive $\phi$ integrals of Eq. \ref{eq:p_split} respectively and exploiting the Gaussian distribution shift invariance, $p_{m}(\Delta T)$ can be expressed as:

\begin{align}\label{eq:p_splitG}
    &p_{m}(\Delta T,Z_{\rm av}) = \int_0^{\tau_{\rm max}^-} p_{\rm g}^-(\tau^-) G(\Delta T - \tau^-;m_{0}(Z_{\rm av}), \sigma_{\rm eff}(Z_{\rm av})) d \tau^- + \int_0^{\tau_{\rm max}^+} p_{\rm g}^+(\tau^+) G(\Delta T - \tau^+;m_{0}(Z_{\rm av}), \sigma_{\rm eff}(Z_{\rm av})) d \tau^+ 
\end{align}

where we made the explicit the separation between the macroscopic geometric term $p_{\rm g}^{\pm}(\tau)$ and the microscopic transport Gaussian kernel. 

Since the average measured  PMT time extension $\langle \Delta T \rangle$ is defined as:
\begin{align} \label{eq:nid_dtav}
&\langle \Delta T \rangle(Z_{\rm av})_m = \int d\Delta T\, \Delta T \,p_{m}(\Delta T,Z_{\rm av}) 
\end{align}
by using the explicit expression of the measured distribution introduced in Eq.~(\ref{eq:p_splitG}),
it can be rewritten as
\begin{align}
\langle \Delta T \rangle(Z_{\rm av})_m
&=
\int d\Delta T \, \Delta T \,
\int_0^{\tau^-_{\rm max}} d\tau^- \,
p_g^-(\tau^-)\,
G(\Delta T-\tau^-;m_0(Z_{\rm av}),\sigma_{\rm eff}(Z_{\rm av})) \nonumber\\
&\quad +
\int d\Delta T \, \Delta T \,
\int_0^{\tau^+_{\rm max}} d\tau^+ \,
p_g^+(\tau^+)\,
G(\Delta T-\tau^+;m_0(Z_{\rm av}),\sigma_{\rm eff}(Z_{\rm av})) .
\end{align}

The integration over $\Delta T$ and $\tau^{\pm}$ can be interchanged since all integrands are positive
and absolutely integrable. Introducing the shifted variables
\begin{equation}
y^{\pm} = \Delta T - \tau^{\pm},
\end{equation}
the two contributions become
\begin{align}
\langle \Delta T \rangle(Z_{\rm av})_m
&=
\int_0^{\tau^-_{\rm max}} d\tau^- \,
p_g^-(\tau^-)
\int dy^- \, (\tau^- + y^-)\,
G(y^-;m_0(Z_{\rm av}),\sigma_{\rm eff}(Z_{\rm av})) \nonumber\\
&\quad +
\int_0^{\tau^+_{\rm max}} d\tau^+ \,
p_g^+(\tau^+)
\int dy^+ \, (\tau^+ + y^+)\,
G(y^+;m_0(Z_{\rm av}),\sigma_{\rm eff}(Z_{\rm av})) .
\end{align}

Each inner integral can be evaluated explicitly by separating the deterministic and stochastic
contributions,
\begin{equation}
\int dy^{\pm}\,(\tau^{\pm}+y^{\pm})\,G(y^{\pm};m_0,\sigma_{\rm eff})
=
\tau^{\pm}
\int dy^{\pm}\,G(y^{\pm};m_0,\sigma_{\rm eff})
+
\int dy^{\pm}\,y^{\pm}G(y^{\pm};m_0,\sigma_{\rm eff}),
\end{equation}
and using the normalization and first moment of the Gaussian kernel,
\begin{equation}
\int dy^{\pm}\,G(y^{\pm};m_0,\sigma_{\rm eff})=1,
\qquad
\int dy^{\pm}\,y^{\pm}G(y^{\pm};m_0,\sigma_{\rm eff})=m_0(Z_{\rm av}).
\end{equation}

This yields to:

\begin{align}
\langle \Delta T \rangle(Z_{\rm av})_m
=\int_{0}^{\tau^-_{\rm max}} p_g^-(\tau^-)
\big( \tau^{-}(\phi) + m_{0} (Z_{\rm av})\big)\, d\tau^-
+ \int_{0}^{\tau^+_{\max}} p_g^+(\tau^+)
\big( \tau^{+}(\phi) + m_{0}(Z_{\rm av}) \big)\, d\tau^+
\end{align}

Since $p_g^-(\tau^-)$ and $p_g^+(\tau^+)$ share the total weight of the angular fan, their integrals sum to unity,
and the expression reduces to:
\begin{align}
    &\langle \Delta T \rangle(Z_{\rm av})_m = m_{0} (Z_{\rm av}) + \frac{\langle \Delta Z \rangle_{g}}{v_l} +\frac{\langle \Delta Z \rangle_{g}}{v_e} 
\end{align}
where $\langle \Delta Z \rangle_{g}$ is the average longitudinal projection of the alphas over the collimated fan. For a narrow vertical fan as the one of our setup, this can be parametrized as $\langle \Delta Z \rangle_{g} = R|\sin\phi_{\rm eff}|$ with $|\sin\phi_{\rm eff}|$ encoding the effective angular weighting. Collecting all contributions and explicitating their dependence on Z$_{\rm av}$ and on the drift field as $v_e = \mu_e E_d$ and $v_l = \mu_ lE_d$, the mean time extension results:

\begin{align}\label{eq:dt_nid_final}
\langle \Delta T \rangle (Z_{\rm av},E_d)_m &= \frac{1}{E_d}\Big[ \left(\frac{1}{\mu_l}+\frac{1}{\mu_e}\right)R|\sin\phi_{\rm eff}| + \left(\frac{1}{\mu_l}-\frac{1}{\mu_e}\right) Z_{\rm av}  \\ \nonumber &+ \frac{\sqrt{(2\ln N_{e}^{\rm eff}) (\sigma^2_{0L_{e}} + \sigma^2_{L_{e} }Z_{\rm av}})}{\mu_e} + \frac{\sqrt{(2\ln N_{l}^{\rm eff}) (\sigma^2_{0L_{l}} + \sigma^2_{L_{l} }Z_{\rm av}})}{\mu_l}  \Big],
\end{align}

where $N_{e/l}^{\rm eff}$ is the effective number of earliest/latest primary ionisation cluster contributing to the arrival times smearing at the track borders, obtained by averaging the angular dependent contributions over the fan geometry. The first contribution of Eq. \ref{eq:dt_nid_final} arises from the geometric projection $\langle \Delta Z \rangle_{g}$ of the extended track along the drift direction, averaged over the fan-like angular acceptance of the collimated source. The second term originates from the possible coexistence of charge carriers with different drift velocities, in case of which the temporal separation between extremes accumulates linearly with the drift distance, i.e. $\tau_0(Z_{\rm av})$. The third and fourth terms encode the stochastic broadening of the arrival times at the track boundaries arising from extreme value statistics applied to the longitudinal diffusion $\sigma_{L_{l/e} }$ and intrinsic spread at the amplification plane $\sigma_{0L_{l/e} }$ of the charge clouds associated with the earliest and latest carriers. The expression obtained above provides a complete quantitative description for the present experimental geometry. In the following subsections, we explicitly separate the single-species (electron drift) and multi-species (negative ion drift) limits in order to make transparent the physical origin of the $\sqrt{Z_{\rm av}}$ and linear $Z_{\rm av}$ scalings used in the main text.

\subsubsection*{Electron Drift Regime (i.e. single specie case)}
In the case of a single specie charge carriers, as in electron drift regime, by construction v$_e$ = v$_l$ = v$_d$, which implies $\tau^+(\phi) = \tau^-(\phi) = \Delta Z(\phi)/v_d$ and therefore

\begin{align}
    m_{\rm 0}^{\rm ED}(Z) & = 2 \sigma_t\sqrt{2 \ln N_{\rm el}^{\rm eff}} \quad \sigma_{\rm eff}^{\rm ED}(Z) =  \sqrt{2}\sigma_t = \sqrt{2}\frac{\sqrt{\sigma^2_{0L_{\rm el}} + \sigma^2_{L_{\rm el} }Z_{\rm av}}}{v_d}
\end{align}
where $\sigma_{0L_{\rm el}}$ is the ED longitudinal diffusion coefficient, $\sigma_{0L_{\rm el}}$ the ED primary clusters smearing at the amplification plane, and $N_{\rm el}^{\rm eff}$ represent the number of ED primary ionisation cluster effectively contributing to the arrival times smearing at the track borders, weighted by the convolution with the geometrical factor. The Gaussian kernel $G(\Delta T - \tau; m_{0}^{\rm ED}, \sigma_{\rm eff}^{\rm ED})$ therefore shifts the purely geometric limit by $m_{0}^{\rm ED}$, while $\sigma_{\rm eff}^{\rm ED}$ sets the intrinsic temporal broadening introduced by longitudinal diffusion, determining the width and local shape of the measured $\Delta T$ distribution. In a single specie ED regime, the expression  for the average measured time extension $\langle \Delta T \rangle^{\rm ED} (Z_{\rm av},E_d)_m$ therefore reduces to:

\begin{align}\label{eq:dt_ed_final}
\langle \Delta T \rangle^{\rm ED} (Z_{\rm av},E_d)_m = \frac{2}{\mu_d E_d }\Big[ R|\sin\phi_{\rm eff}| + \sqrt{(2\ln N_{\rm el}^{\rm eff}) (\sigma^2_{0L_{\rm el}} + \sigma^2_{L_{\rm el} }Z_{\rm av}})  \Big],
\end{align}
displaying a characteristic $\sqrt{Z_{\rm av}}$ arising from the stochastic diffusion contribution to extremes, in addition to the geometrical $R|\sin\phi_{\rm eff}|$ term.

\begin{figure}[!t] 
 \includegraphics[width=0.45\linewidth]{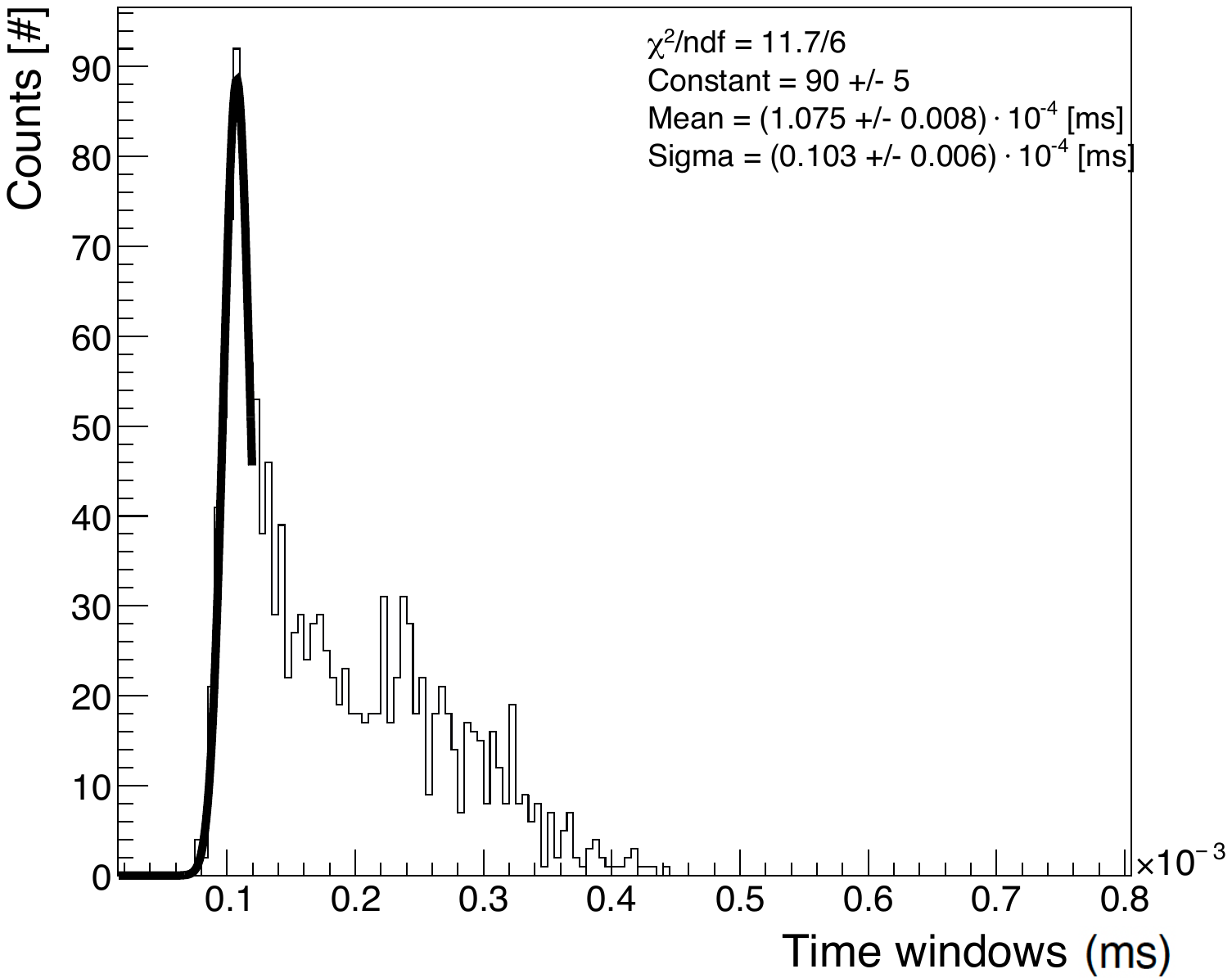} \includegraphics[width=0.4\linewidth]{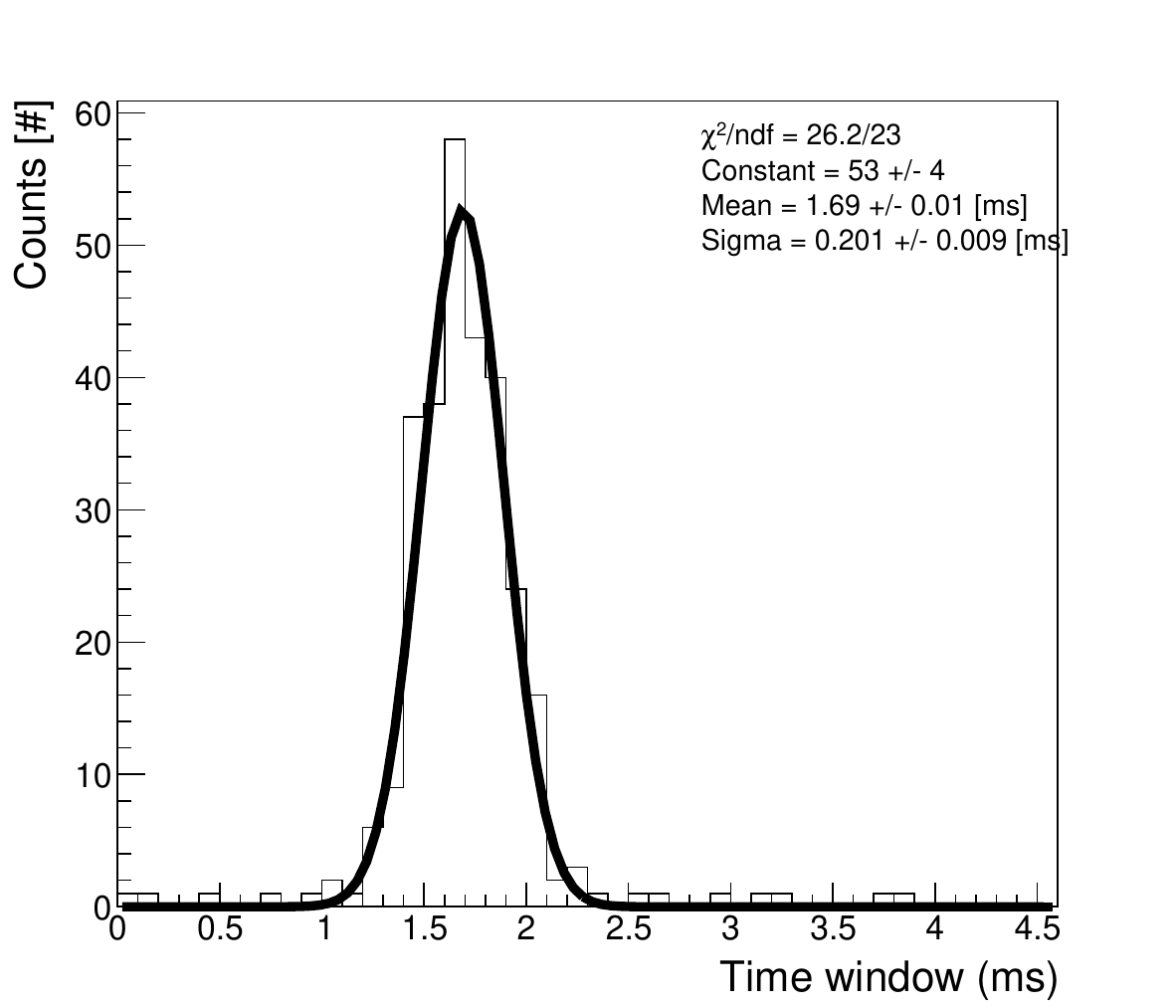}
	\caption{Representative example of measured time extensions $\Delta T$ distributions at 650 mbar, Z$_{\rm av}$=3.5 cm and E$_d$=300 V/cm for ED (left) and NID (right), with a Gaussian fit superimposed as discussed in the text.}
	\label{fig:time_dist}
\end{figure}

The illustrated framework appears evident in the measured distribution of the ED $\Delta T$ at 650 mbar, Z$_{\rm av}$ = 3.5 cm drift distance and E$_d$= 300 V/cm in Fig. \ref{fig:time_dist}. 
The large probability densities at small $\Delta T$ and a monotonically decreasing tail vanishing at the geometric endpoint $\Delta T_{\max}^{\rm geom}$ set by the collimation cone reflect the Jacobian transformation in Eq. \ref{eq:geom}.
The peak at low values arises from tracks emitted nearly parallel to the GEM plane ($|\phi|\simeq0$), for which the geometrical separation $\Delta Z(\phi)$ becomes so small that the earliest–latest arrival time difference is entirely governed by the intrinsic diffusion kernel, producing a peak centered around the microscopic scale $m_0$. The upper cutoff of the distribution is fixed by the convolution of the Gaussian kernel with the macroscopic acceptance of the collimator slit, extending the purely geometric limit to $\Delta T^{\rm meas}_{\max}= \Delta T^{\rm geom}_{\max} + m_{0}^{\rm ED}$. A Gaussian fit to the ED $\Delta T$ peak at low values returns $ m^{\rm ED}_{0} = 107.6 \pm 0.8$ ns, consistent $R |\sin \phi_{\max}| \sim$  1 cm expected from the thin vertical fan collimation of our setup, using $v_d = 0.0032$ cm/ns and $\sigma_{0L_{\rm el}}$ = 0.0165 cm/$\sqrt{\rm cm}$ from Garfield++ simulation, assuming $\sigma_{0L_{\rm el}} \simeq$ 0.04 cm from Ref. ~[S3] and extrapolating $\Delta T^{\rm meas}_{\max} \sim$ 450 ns from Fig.\ref{fig:time_dist}.

\subsubsection*{Negative Ion Drift Regime (i.e. multi-species case)}
The NID picture results further enriched by the possible presence of multiple negative ion carriers with different drift velocities, as already observed in CS$_2$ mixtures ~[S4]  and pure SF$_6$ ~[S5, S6]. Concerning in particular the discussion about water vapor contamination at 20 Torr operation reported in ~[S5], and considering the humidity level of $\sim$0.1$\%$ measured in our setup, the presence of SOF$^{-}_4$ and F$^{-}$(HF)$_2$ negative minority ion clusters in our measurements appears reasonable. This implies by construction v$_e \neq$ v$_l$. Since NID drift with velocities of O(cm/ms) ~[S7] and the shortest drift distance acquired is 2.5 cm, a mobility difference of $\le \sim$ O(1) $\%$ between faster and slower charge carriers would be required for the EVT broadening to overcome the deterministic drift time splitting, a regime far from the O(10$\%$) faster mobilities observed in minority carriers  ~[S5, S8]. For this reason, in the NID regime we can assume that the faster carrier with velocity v$_f$ will always arrive before the slower one drifting at v$_s$, regardless of the geometry of the track, with the identifications $v_e \equiv v_f$ and $v_l \equiv v_s$. In addition, since no physical mechanism is expected to generate substantially different longitudinal diffusion widths for the two anion species in this mixture, we adopt the approximation $\sigma_{\rm diff,L}^{f} (Z) \simeq \sigma_{\rm diff,L}^{s} (Z) \equiv \sigma_{\rm diff,L}^{\rm NID} (Z)$.

Therefore, assuming in the NID regime the presence of two (or more) species of anions (consisting of $N_s$ and $N_f$ cluster respectively), with drift with velocities $v_{\text{f}}$ and $v_{\text{s}}$, with $v_{\text{f}} > v_{\text{s}}$, but identical diffusion properties, the NID Gaussian kernel mean and sigmas of $ p_{\rm meas}^{\rm NID}(\Delta T)$ in Eq.(\ref{eq:p_splitG}) can be written as:
\begin{align}\label{eq:m0_nid}
    m_{0}^{\rm NID}(Z) &=   Z \Big ( \frac{1}{v_s} - \frac{1}{v_f}\Big ) +  \sigma_{\rm diff,L}^{\rm NID} \Big(\frac{\sqrt{2\ln N_{s}}}{v_s} + \frac{\sqrt{2\ln N_{f}}}{v_f}  \Big), \\ \nonumber
    \sigma_{\rm eff}^{\rm NID}(Z) &= \sigma_{\rm diff,L}^{\rm NID} \sqrt{\frac{1}{v_s^2} + \frac{1}{v_f^2}} 
\end{align}

with $\tau^-(\phi) = \Delta Z(\phi)/v_f$ and $\tau^+(\phi) = \Delta Z(\phi)/v_s$.

The measured NID $\Delta T$ distributions exhibit a Gaussian shape over the full range of drift distances and drift fields tested, with an average displaying a clear and systematic linear trend on $Z_{\rm av}$ as expected from Eq. (\ref{eq:m0_nid}). Such observed dependence cannot be accounted for by diffusion, amplification smearing, or geometrical effects alone, and therefore this behaviour provides direct experimental evidence of the coexistence of at least two anion populations in the NID gas mixture with different velocities. Here the slow component is expected to correspond to the dominant $\mathrm{SF_6^-}$ negative ion, while the faster carrier represents the most mobile among the minor anion species, which we do not attempt to identify here. Within the assumption illustrated above, the NID mean time extension can be expressed as:

\begin{equation}\label{eq:dt_nid_final}
\langle \Delta T \rangle^{\rm NID} (Z_{\rm av},E_d) = \frac{1}{E_d}\Big( \alpha + \beta Z_{\rm av} + \gamma \sqrt{\sigma_{0L}^2 + \sigma_L^2 Z_{\rm av}} \Big),
\end{equation}

having defined 
\begin{align}
     \alpha &= \left(\frac{1}{\mu_s}+\frac{1}{\mu_f}\right)R|\sin\phi_{\rm eff}| \\  \nonumber
     \beta &=\left(\frac{1}{\mu_s}-\frac{1}{\mu_f}\right), \\ \nonumber
     \gamma &= \frac{\sqrt{2\ln N_{s}^{\rm eff}}}{\mu_s} +  \frac{\sqrt{2\ln N_{f}^{\rm eff}}}{\mu_f}
\end{align}

where $N_{s/f}^{\rm eff}$ is the effective number of fast/slow primary ionisation cluster contributing to the arrival times smearing at the track borders, obtained by averaging the angular dependent contributions over the source collimation fan geometry.

For the NID dataset presented in this paper, Eq.\ref{eq:dt_nid_final} can be further simplified considering the relative weights $\beta$ and $\gamma$ of the $\propto Z_{\rm av}$ and $\propto \sqrt{Z_{\rm av}}$ terms.  Since for thermal diffusion $\sigma_L \simeq$ 0.01 cm/$\sqrt{\rm cm}$ and  $\mu_{s,f} \simeq$ O(1) cm$^2$ V$^{-1}$ s$^{-1}$ are expected, assuming a similar contribution within the GEMs in the ED and NID regime $\sigma_{0L}\sim 0.04$ cm ~[S3], $\gamma \ll \beta$, so that the $\sqrt{Z_{\rm av}}$ term can be neglected for all Z$_{\rm av}$. In addition, since the alpha track produces in our setup O(10$^5$) primary clusters N, even for a minority cluster population of O(10$\%$) of the majority charge carriers, $\sqrt{2\ln N_{s}^{\rm eff}} \sim \sqrt{2\ln N_{f}^{\rm eff}} \sim \sqrt{2\ln N^{\rm eff}}$, and therefore we can effectively parameterise the fitted $\Delta T$ distribution Gaussian means in our dataset as a function of drift distance and field as:

\begin{equation}\label{eq:dt_p0}
\langle \Delta T \rangle_{} (Z_{\rm av},E_d)_{m} = \frac{1}{E_d}\Big( p_0 + p_1Z_{\rm av} \Big),
\end{equation}

with

\begin{align}
    p_0 &= \left(\frac{1}{\mu_s}+\frac{1}{\mu_f}\right) \left(R|\sin\phi_{\rm eff}| +  \sigma_{0L}\sqrt{2\ln N_{t}^{\rm eff}}\right) \\ \nonumber
    p_1 &=\left(\frac{1}{\mu_s}-\frac{1}{\mu_f}\right)
\end{align}

which is the form adopted in the main text to describe the measured NID data.

\section*{References}
{\small
\begin{list}{}{\leftmargin=0.9cm \itemindent=-0.9cm \itemsep=1pt \parsep=0pt \topsep=2pt}

\item[{[S1]}] E. A. Mason and E. W. McDaniel, \textit{Transport Properties of Ions in Gases} (Wiley, New York, 1988).

\item[{[S2]}] S. N. Majumdar, A. Pal, and G. Schehr, Phys. Rep. \textbf{840}, 1 (2020).

\item[{[S3]}] L. Zappaterra, \textit{Studies of the performance of the CYGNO experiment in low-energy nuclear recoils detection}, Ph.D. thesis, Universit\`a degli Studi di Roma ``La Sapienza'' and CERN (2024).

\item[{[S4]}] D. P. Snowden-Ifft and J. L. Gauvreau, Rev. Sci. Instrum. \textbf{84}, 053304 (2013).

\item[{[S5]}] N. S. Phan, R. Lafler, R. J. Lauer, E. R. Lee, D. Loomba, J. A. Matthews, and E. H. Miller, J. Instrum. \textbf{12}, P02012 (2017).

\item[{[S6]}] T. Ikeda, T. Shimada, H. Ishiura, K. D. Nakamura, T. Nakamura, and K. Miuchi, J. Instrum. \textbf{15}, P07015 (2020).

\item[{[S7]}] E. Baracchini, G. Cavoto, G. Mazzitelli, F. Murtas, F. Renga, and S. Tomassini, J. Instrum. \textbf{13}, P04022 (2018).

\item[{[S8]}] D. P. Snowden-Ifft, Rev. Sci. Instrum. \textbf{85}, 013303 (2014).

\end{list}
}

\end{document}